\documentclass[a4paper,10pt,aps,pre,twocolumn,superscriptaddress,amsmath,amssymb,nofootinbib]{revtex4}
\usepackage{graphicx,color,soul}
\usepackage[colorlinks=true,linkcolor=blue]{hyperref}%

\usepackage[dvipsnames]{xcolor}
\colorlet{Mycolor1}{green!10!orange!90!}

\begin{document}

\title{Coupling between criticality and gelation in ``sticky'' spheres: A structural analysis}

\author{David Richard}
\affiliation{Institut f\"ur Physik, Johannes Gutenberg-Universit\"at Mainz, Staudingerweg 7-9, 55128 Mainz, Germany}
\author{James Hallett}
\affiliation{HH Wills Physics Laboratory, Tyndall Avenue, Bristol BS8 1TL, UK}
\author{Thomas Speck}
\affiliation{Institut f\"ur Physik, Johannes Gutenberg-Universit\"at Mainz, Staudingerweg 7-9, 55128 Mainz, Germany}
\author{C. Patrick Royall}
\affiliation{HH Wills Physics Laboratory, Tyndall Avenue, Bristol BS8 1TL, UK}
\affiliation{School of Chemistry, University of Bristol, Cantock's Close, Bristol, UK}
\affiliation{Centre for Nanoscience and Quantum Information, Tyndall Avenue, Bristol, UK}

\begin{abstract}
  We combine experiments and simulations to study the link between criticality and gelation in sticky spheres. We employ confocal microscopy to image colloid-polymer mixtures, and Monte Carlo simulations of the square-well (SW) potential as a reference model. To this end, we map our experimental samples onto the SW model. We find excellent structural agreement between experiments and simulations, both for locally favored structures at the single particle level and large-scale fluctuations at criticality. We follow in detail the rapid structural change of the critical fluid when approaching the gas-liquid binodal and highlight the role of critical density fluctuations for this structural crossover. Our results link the arrested spinodal decomposition to long-lived energetically favored structures, which grow even away from the binodal due to the critical scaling of the bulk correlation length and static susceptibility. 
\end{abstract}

\maketitle

% ---- introduction ----

\section{Introduction}

Understanding how an amorphous system becomes dynamically arrested upon compression or cooling is a long-standing challenge in statistical physics. Such amorphous solids encompass states of matter such as glasses, films, plastics, and gels, among others. Despite the fact that these systems are of technological importance and have received a lot of attention in the literature~\cite{debenedetti2001supercooled,berthier2011theoretical}, the microscopic mechanism responsible for macroscopic arrest remains elusive. For instance, the question whether the glass transition can be explained in the context of a thermodynamic or structural phase transition is still debated~\cite{biroli2013perspective,berthier2016,turci2017prx,royall2015role,royall2017}. Interestingly, gels, in contrast to glasses, can have very sparse spatial structural arrangements, often described by a percolating network of bonded particles whose degree of dilution can in principle be unbounded~\cite{manley2005glasslike,griffiths2017local}. These networks can result from cross-linking polymer chains or from physical bonds. The latter are caused, \emph{e.g.}, by depletion attractions~\cite{poon2002physics} and for some time have served as a well controlled model system to study glasses and gels~\cite{van1991dynamic,de2001direct,poon2002physics,cipeletti2005,zaccarelli2007colloidal,meng2010free,royall2008direct,taylor2012temperature,pinchaipat2017}.

When decreasing the range of attraction between particles, the gas-liquid coexistence becomes metastable with respect to fluid-solid coexistence, with a metastable critical point~\cite{haxton2015crystallization,largo2008vanishing}. In the limit where the attraction range becomes smaller than roughly 10\% of the diameter of the particles, the shape of the attractive part of the pair-wise potential becomes irrelevant~\cite{noro2000extended,largo2008vanishing}. This gave rise to the extended law of corresponding states~\cite{noro2000extended}, which can also explain some aspects of the phase behavior of protein solutions~\cite{stradner2004equilibrium,katsonis2006corresponding,platten2015extended}. Experimentally, such a short-ranged attraction can be probed using colloid-polymer mixtures~\cite{poon2002physics}, so-called ``sticky'' spheres. Here, the polymers play the role of depletants for the large colloidal particles, where the radius of gyration of the polymer chains sets the range of attraction.

In the context of understanding the gelation of particles with short-ranged attraction, much work has been done during the last 20 years~\cite{grant1993volume,verduin1995phase,poon2002,cipeletti2005,zaccarelli2007,manley2005glasslike,lu2008gelation,royall2008direct,laurati2009,eberle2011dynamical,pandey2013,royall2015probing,royall2018vitri}. Several scenarios have been proposed, including diffusion-limited cluster aggregation~\cite{bibette1992kinetically,kroy2004cluster}, phase separation~\cite{manley2005glasslike,lu2008gelation}, and percolation~\cite{verduin1995phase,valadez2013dynamical,kohl2016directed}; and these can couple in different regions of the phase diagram~\cite{griffiths2017local}. Some of this work links the gelation to an arrested spinodal decomposition~\cite{verhaegh1997,manley2005glasslike,lu2008gelation,royall2008direct} and thus, a direct consequence of the underlying gas-liquid coexistence. A possible microscopic mechanism for the arrest was proposed with the presence, at gelation, of clusters that minimise the local potential energy~\cite{royall2008direct}. Furthermore, it has been shown that gelation in sticky spheres offers a clear dynamical signature in comparison with a hard-sphere glass, where authors found a quasi-discontinuous increase in the relaxation time of the fluid for various packing fractions~\cite{royall2018vitri}. These dynamical transitions were found to be located at the gas-liquid binodal, even for a very dense gel, with a packing fraction exceeding the freezing point of the hard-sphere fluid~\cite{royall2018vitri}. Other work has suggested that the gelation line is located before the phase separation and extends at higher densities toward the location of the attractive glass~\cite{pham2002multiple}. A more recent numerical study~\cite{valadez2013dynamical} of the adhesive hard-sphere model~\cite{baxter1968percus,buzzaccaro2007sticky} has linked the experimental gelation line to the mean-field rigidity transition introduced by He and Thorpe~\cite{he1985elastic} in the context of random networks. More recently, a connection between directed percolation~\cite{hinrichsen2000non} as an equilibrium pre-structural transition to gelation has been proposed~\cite{kohl2016directed}.

The suggested relationship between the liquid-gas phase separation line~\cite{verhaegh1997,lu2008gelation,royall2018vitri} and gelation implies that critical fluctuations may influence the gelation process in the vicinity of the critical isochore. While criticality has been studied in colloid-polymer systems~\cite{bodnar1996,royall2007measuring}, this work has tended to focus on systems where the interaction range is long enough that the system exhibits a stable colloidal liquid phase. In colloid-polymer mixtures, this corresponds to a polymer-colloid size ratio of around 0.3~\cite{lekkerkerker1992,poon2002physics}. For such systems, the colloidal liquid is not dense enough to arrest, and gelation is only achieved upon quenching with a much stronger attraction strength than that requited for criticality~\cite{zhang2013}. However, in the sticky sphere limit, immediately upon quenching through the binodal, the density of the colloidal ``liquid'' is sufficient that the system undergoes dynamical arrest~\cite{royall2018vitri}. Under these circumstances, gelation may couple to critical fluctuations and this forms the subject of our study. 

%% ---- model ----

\section{Methods}
\label{methods}

\subsection{Experiments}

We employ confocal microscopy and particle tracking to resolve the positions of the colloidal particles. Colloid-polymer mixtures are composed of polystyrene (PS) polymer chains and sterically stabilized polymethyl methacrylate (PMMA) spheres with diameter $\sigma=2950\,\text{nm}$ and polydispersity $\Delta=5\%$, determined via scanning electronic microscopy (SEM). We use rhodamine as fluorescent label. Polystyrene has a molecular weight $M_w=1.3\times10^6$ corresponding to an effective radius of gyration $R_g=35\,\text{nm}$ under $\theta$ conditions. We use a solvent mixture of cis-decalin and cyclohexyl bromide which is density and refractive index-matched. We additionally screen electrostatic interactions using 4 mMol of tetrabutyl ammonium bromide salt. From previous work~\cite{taylor2012temperature}, we estimate $R_g\simeq50\,\text{nm}$ at room temperature. This leads to a polymer-colloid size ratio $q\simeq0.03$ approaching the sticky sphere limit. We then seal each sample into a borosilicate glass capillary with epoxy resin.

Samples are imaged by confocal microscopy using a Leica SP8. We image different parts of the suspension at least $15\,\mu\text{m}$ from the wall. To extract the colloidal particle positions we employ three dimensional particle tracking using the difference of Gaussian method. This tracking is performed using the package ''Colloid'' developed by Leocmach \emph{et al.}~\cite{leocmach2013novel}. From the tracking we estimate the sample packing fraction as $\phi \simeq \pi N\sigma^3/(6V)$, with $N$ the number of tracked particles and $V$ the volume of the sample. It is well known that sterically stabilized PMMA colloidal particles can exhibit swelling and unscreened electrostatic interactions~\cite{poon2012measuring,royall2013search} which can lead to an intrinsic softness modeled by an effective diameter $\sigma_{\text{eff}}>\sigma$. To estimate $\sigma_{\text{eff}}$ under our experimental conditions, we apply a method similar to the one used in Ref.~\cite{turci2016crystallisation}. In this study, the authors matched the pair correlation function $g(r)$ of dense hard spheres with the known Percus-Yevick expression by varying the effective diameter $\sigma_{\text{eff}}$. In our work, we extend this mapping with matching the $g(r)$ of the colloid-polymer mixture with an attractive square-well fluid, which we adopt throughout this study as a reference system for our mixtures. In addition, our mapping allows us to determine the systematic tracking errors of the colloidal positions, responsible for the broadness of $g(r)$ at contact~\cite{royall2007measuring}. More details can be found in the appendix~\ref{sigma}. We find $\sigma_\text{eff}=3100 \ \text{nm}$, which corresponds to an effective diameter that is $\sim5\%$ larger than for the SEM estimation.

\subsection{Computer simulations}

We study the behavior of the square-well (SW) model serving as a reference model for our colloid-polymer mixtures. We perform standard Monte Carlo simulations in the NVT ensemble employing local moves. The system is composed of $N=5000$ particles with diameters drawn from a Gaussian distribution with polydispersity $\Delta=5\%$. The interaction potential between two particles $i$ and $j$ is
\begin{equation}
V(r)=\begin{cases}
\infty & \text{if $r \le \sigma_{ij}$}\\
-V_{\text{SW}} & \text{if $\sigma_{ij} < r <\sigma_{ij}+\delta$}\\
0 & \text{if $r \ge \sigma_{ij}+\delta $},
\end{cases}
\end{equation}
where $\sigma_{ij}=(\sigma_i+\sigma_j)/2$, $\sigma_i$ and $\sigma_j$ being the diameter of particle $i$ and $j$ respectively. The attraction range $\delta$ is set by the polymers and fixed to $0.03\sigma$ and independ of $\sigma_{ij}$. If not mentioned otherwise, simulations are performed in a cubic box. Additionally, we use a slab geometry to compute the coexistence packing fraction between the gas and liquid phase with box lengths: $L_x=L_y=L_z/2$. We determine the binodal by fitting the density profile along the gas-liquid coexistence through,
\begin{equation}
\phi(z)=\frac{\phi_l-\phi_g}{2}+\frac{\phi_l-\phi_g}{2}\tanh\Big(\frac{z-z_0}{2w}\Big).
\label{eq:profile}
\end{equation}
Here, $\phi_g$ and $\phi_l$ denote the gas and liquid coexistence packing fractions, and $z_0$ and $w$ the interface position and width, respectively. Since for this polydispersity the liquid is still metastable with respect to the crystal, we check that all liquid slabs remained in the liquid phase. All distances and energies are expressed in units of $\sigma$ and $k_BT$, $k_B$ being the Boltzmann constant. Throughout, we denote the dimensionless attraction strength by $V=V_\text{SW}/k_BT$.

\subsection{Mapping procedure}

The attraction strength between colloidal particles is controlled by the polymer concentration $c_p$. However, it is challenging to determine $c_p$ precisely enough to map it directly. Instead, we map every sample individually through matching the experimentally measured total correlation function $h_{\text{exp}}(r)=g_{\text{exp}}(r)-1$ and distribution $P_{\text{exp}}(n)$ of bond number $n$ to the SW fluid~\cite{royall2007measuring,lu2008gelation}, yielding an effective attraction strength $V$ for every experimental sample.

\begin{figure}[t]
  \includegraphics[scale=1.0]{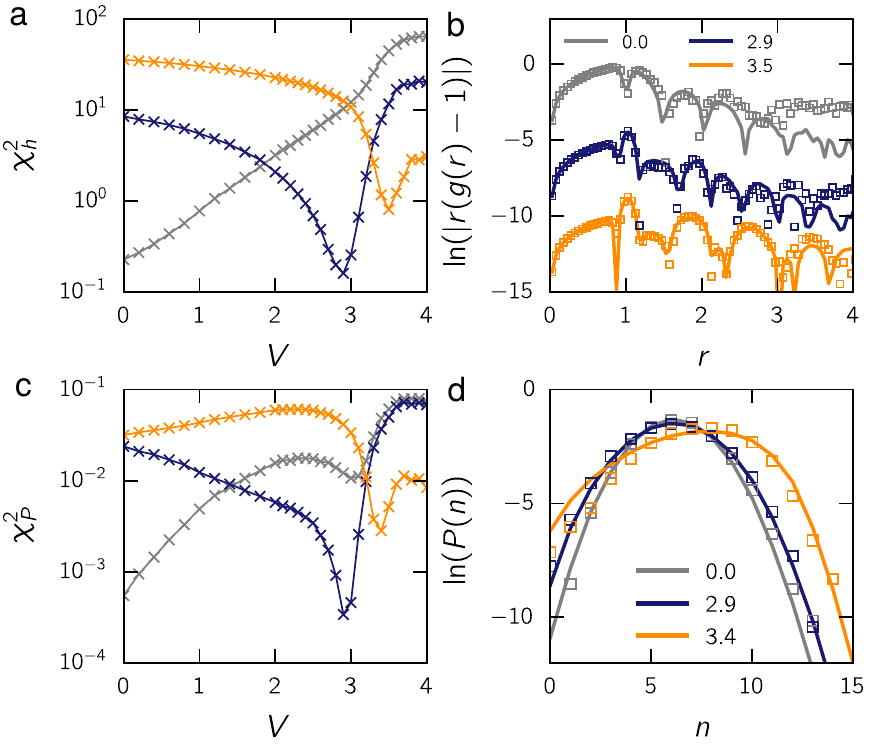}
  \caption{\textbf{Mapping onto the square-well model.} (a,b)~Evolution of the least squares $\chi^2$ as a function of the attraction strength $V$ for the total correlation function $rh(r)$ (a) and bond distribution $P(n)$ (b) of three samples. (c,d) Comparison of $rh(r)$ (c) and $P(n)$ (d) of the experimental samples (empty symbols) with the SW model for matched $V$ (lines). The color code distinguishes a hard-sphere fluid (gray) from a critical (blue) and gel (orange) sample.}
  \label{fig:mapping}
\end{figure}

Specifically, we compute $rh_{\text{sim}}(r)$ and  $P_{\text{sim}}(n)$ on a grid in the ($\phi$, $V$) plane around the critical point. We evaluate $P(n)$ by constructing a bond network using a Voronoi decomposition considering only direct Voronoi neighbors~\cite{brostow1978construction,malins2013identification} with the bond distance cutoff $r_c$ set to $1.5\sigma$. Following the work of Largo~\textit{et al.}, we estimate the location of the critical point to be at $\phi_c\simeq0.275$ and $V_c\simeq3.22$ for zero polydispersity~\cite{largo2008vanishing}. We shall show later on that this value is very close to the critical point of our model with $\Delta=5\%$. We use packing fractions ranging from $\phi=0.05$ to $0.5$ with an interval $\Delta\phi=0.025$ and $V$ values ranging from $0$ to $4$ with $\Delta V=0.2$ for $0<V<2$ and $\Delta V=0.1$ for $2<V<4$. Overall, we end up with a grid of $608$ state points. Beyond $V\simeq3.2$, the fluid crosses the binodal and starts to form a gel. Thus, structural observables such as $h(r)$ and $P(n)$ might evolve due to aging. We fix for every state point a MC relaxation time of $10^6$ steps before we compute any observables. Another $10^6$ steps is used to compute observables. We then pick for each sample the two numerical packing fractions  $\phi_{-}$ and $\phi_{+}$, which encompass our sample density. We then compute for each $V$ value $h_{\text{sim}}$ and $P_{\text{sim}}$ as a linear combination of $\phi_{-}$ and $\phi_{+}$. We finally compute as a ``goodness'' parameter for our matching procedure the least squares
\begin{equation}
  \chi^2_h = \sum_i [ r_i(h_{\text{exp}}(r_i) - h_{\text{sim}}(r_i))]^2
\end{equation}
and
\begin{equation}
  \chi^2_P = \sum_i [P_{\text{exp}}(n_i) - P_{\text{sim}}(n_i)]^2.
\end{equation}
In practice, $\chi^2_h$ is computed for $r<4\sigma$, while an additional Gaussian ($\Delta=5\%$) noise is added on numerical positions to mimic particle tracking errors~\cite{royall2007measuring}, see appendix~\ref{sigma}. The global minima of $\chi^2_h$ and $\chi^2_P$ give us two independent evaluations of $V_{\text{exp}}$. We then assign for each sample the mean of those two, $V_{\text{exp}}=[V_h+V_P]/2$, whereas $(|V_h-V_P|)/2$ serves as an estimation for errors.

In Fig.~\ref{fig:mapping}(a) and (b), we present $\chi^2_h$ and $\chi^2_P$ as functions of $V$ for three samples picked along the critical isochore. From a hard sphere fluid (gray curves) to the gel (red curves), we always observe a global minimum for $\chi^2$ and find a good agreement between $V_h$ and $V_P$. Results of the SW model are shown in Fig~\ref{fig:mapping}(c,d) and compared to experimental data. The correspondence between experiments and simulations is excellent, we observe only small discrepancies in the bond-number distribution for the gel sample for small bond numbers $n$, see Fig~\ref{fig:mapping}(d). These deviations are not surprising since we do not include the effect of aging in our mapping method.

%% ---- Results ----

\section{Results}
\label{results}

\subsection{Phase diagram}
\label{phase}

\begin{figure*}[t]
  \includegraphics[scale=1.0]{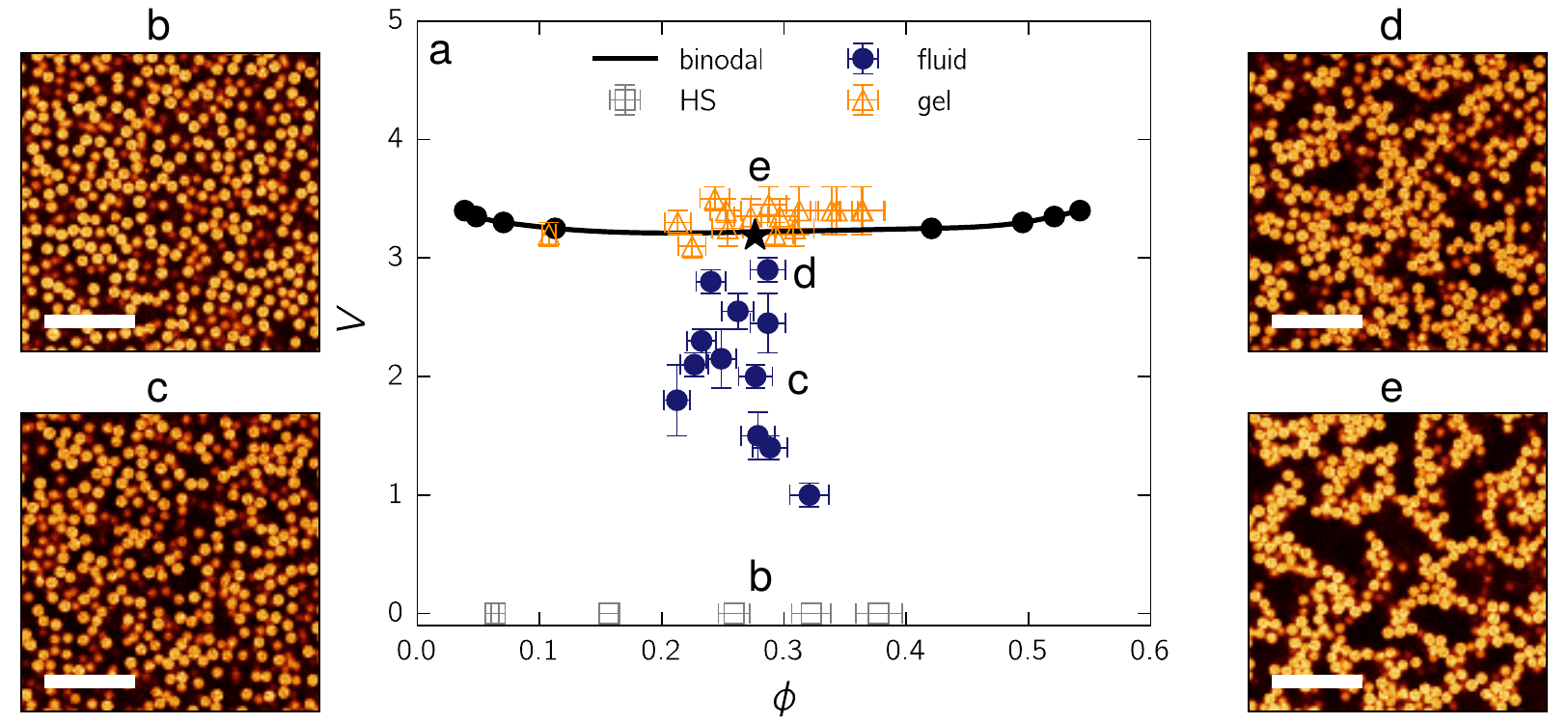}
  \caption{\textbf{Phase diagram.}~(a) Phase diagram of the square-well with $\delta=3\%$ and polydispersity $\Delta=5\%$. Black points indicate the gas-liquid coexistence and the black star indicates the critical point. The black solid line is a guide to the eye for the two phase boundary. Gray, blue, and orange squares are, respectively, hard-sphere fluid, colloid-polymer mixtures, and arrested gel samples. Confocal images are displayed in a, b, c and d: (b) Hard sphere fluid without polymers ($V\simeq0.0$); (c) Equilibrium fluid ($V\simeq2.0$); (d) Fluid close to criticality ($V\simeq2.9$) and (e) Gel phase ($V\simeq3.2$). The scale bars correspond to $20 \ \mu m$.
} 
  \label{fig:phase}
\end{figure*}

We start by discussing the phase behavior of our system in Fig.~\ref{fig:phase}. We choose to present our data in the $\phi$-$V$ plane. The gas-liquid binodal of our reference system is indicated by the black solid line. The location of the critical point is indicated by a black star and was determined via the block distribution functions method~\cite{binder1981finite}, we shall come back to this point later. From our mapping procedure, we can place each sample at a given $\phi$ and $V$. We distinguish an ergodic fluid in contrast to a gel by respectively blue and red square symbols. We also indicate samples without polymers, \emph{i.e.} hard spheres, in gray. The typical gel sample shows a clear dynamical arrest with a relaxation time beyond $100 \tau_B$ (Supplementary Video) with $\tau_B$  being the Brownian time. This is consistent with a previous dynamical experimental study for a very similar system~\cite{royall2018vitri}.

The main observation here is that arrested samples are lying close to the binodal for a wide range of densities ($0.1<\phi<0.4$). That is to say, all samples which are gels always map onto state points of the reference system which are on or above the binodal. This means that the radial distribution function and bond distribution of a gel sample never correspond to an ergodic fluid. This is in agreement with previous numerical and experimental studies, which associate the gelation to the location of the spinodal and thus a direct consequence of equilibrium properties of the gas-liquid coexistence~\cite{manley2005glasslike,lu2008gelation,royall2008direct,royall2018vitri}. For short-ranged attractive systems, the binodal and spinodal are located very close together, such that in experiments it is hard to distinguish them~\cite{royall2018vitri}. The highest attraction strength at which the sample remains ergodic is found at $V\simeq2.9$. Additionally, in Fig.~\ref{fig:phase}, we show images from confocal microscopy along the critical isochore $\phi\simeq\phi_c$. We observe a rather continuous change of structural behavior from hard-sphere fluid to gel. At $V=0$ [Fig.~\ref{fig:phase}(b)] the spatial distribution of colloids is homogeneous at length scales larger than the particle size. When increasing the concentration of polymers, \emph{i.e.} increasing $V$, clusters start to form with lower density in between [Fig.~\ref{fig:phase}(c-d)]. Finally, the suspension shows an arrested spinodal decomposition, which is revealed in Fig.~\ref{fig:phase}(e) by large assemblies rich in colloidal particles and depleted zones without any particles.

\subsection{Critical density fluctuations}
\label{criticality}

\begin{figure}[t]
  \includegraphics[scale=1.0]{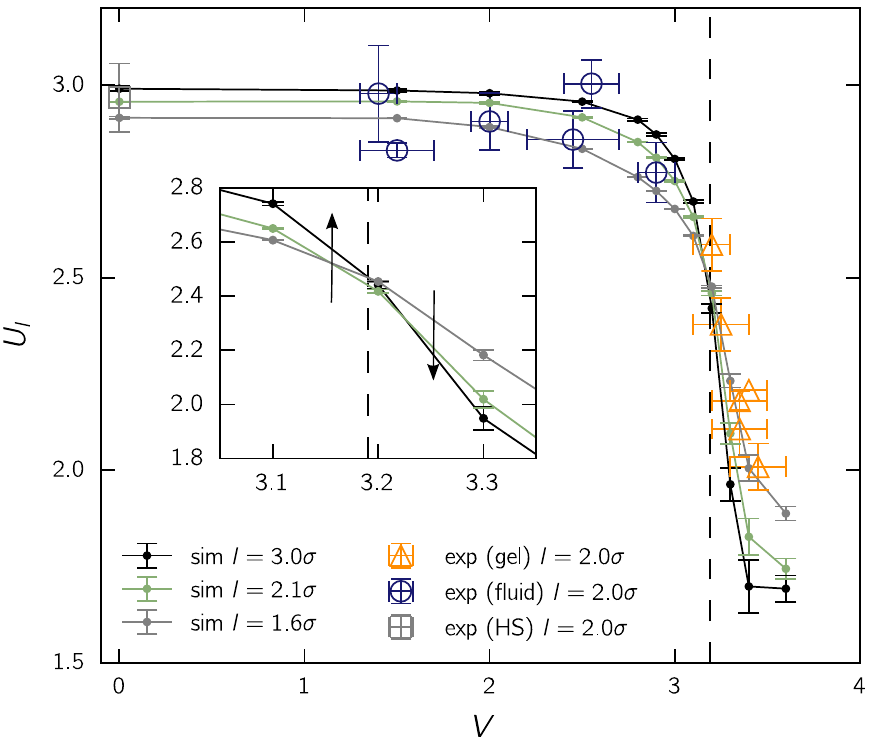}
  \caption{\textbf{Critical point.} Evolution of the Binder cumulant $U_l$ as a function of the attraction strength $V$ for three values of $l$. Solid lines and open symbols indicate simulation and experimental data, respectively. The hard sphere, ergodic fluid, and arrested gel samples are distinguished by gray squares, blue circles and orange triangles respectively. Inset: zoom into the crossing region for simulation data.}
  \label{fig:cumulant}
\end{figure}

In section~\ref{phase}, we have shown that the gelation line is located close to the binodal and that dense domains of colloidal particles grow progressively as a function of the polymer concentration. We now demonstrate how this change of behavior can be directly linked to the fluctuations present in the context of criticality. To quantify the spatial evolution of the density we use the block distribution functions method~\cite{binder1981finite}, which provides the location of the critical point $V_c$~\cite{watanabe2012phase} and to some extent the isothermal susceptibility $\chi$ of the fluid~\cite{rovere1993simulation,sengupta2000elastic}. The procedure is as follows: We divide our system into a series of cubic subcells of dimension $l=L/b$, with $b$ being an integer. The global density is defined as $\overline{\rho} = \frac{1}{n_b}\sum_i \rho_i$, where $n_b=b^3$ is the number of subcells and $\rho_i$ is the local density in subcell $i$. We then extract the second and forth moment of the density $\langle m_2 \rangle$ and $\langle m_4 \rangle$ computed as $\langle m_x \rangle_l = \frac{1}{n_b}\sum_i (\rho_i-\overline{\rho})^x$. We can finally define the Binder cumulant through
\begin{equation}
  U_l = \frac{\langle m_4 \rangle_{l}}{\langle m_2 \rangle_{l}^2}.
\end{equation}
One of the main properties of $U_l$ is its size invariance with respect to $l$ at the critical point~\cite{binder1981finite}. This allows us to determine unambiguously and accurately the critical attraction strength $V_c$ in computer simulations. In more detail, we compute $U_l$ for various attraction strengths $V$ and subcell lengths $l$. To this end, we sample $U_l$ for each $V$ values at a fixed packing fraction $\phi\simeq\phi_c=0.275$ using an extra $10^8$ MC steps. In Fig.~\ref{fig:cumulant}, we show the evolution of $U_l$ as a function of $V$. We observe a crossing for $V>3$, which can be resolved more accurately as shown in the inset of Fig.~\ref{fig:cumulant}(a). We find $V_c=3.19(3)$ as the final value for the critical point, which is close to the value determined by Largo \textit{et al.}~\cite{largo2008vanishing} for zero polydispersity. Additionally, in Fig.~\ref{fig:cumulant} we show experimental results for $U_l$ at a fixed subcell length, $l=2\sigma$. Although the experimental data suffer from a lack of statistic in comparison with simulation data, we observe an overall good agreement. More precisely, gel samples indicated by red squares are, within the errors, located either at the cumulant crossing or at larger attraction strengths.

\begin{figure}[t]
  \includegraphics[width=\linewidth]{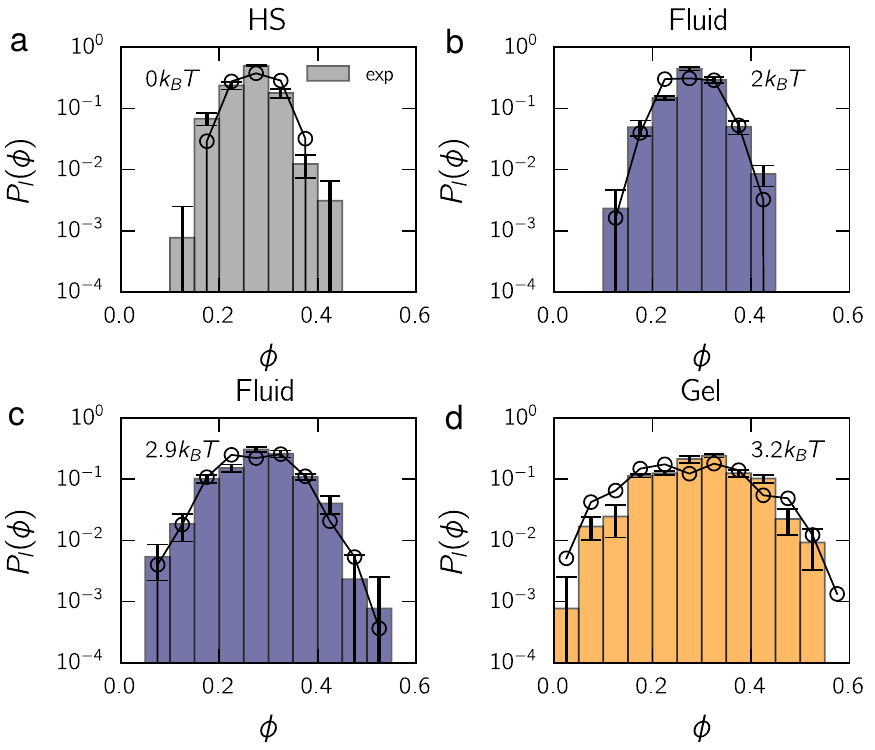}
  \caption{\textbf{Density fluctuations.} Packing fraction distribution $P_l(\phi)$ for various samples and matched simulations along the critical path for subcell length $l\simeq3\sigma$. Colored histograms and empty black circles are respectively experimental and simulation distributions. The effective attraction depth $V$ for each sample is indicated in figures.}
  \label{fig:distribution}
\end{figure}

In practice, the change of $U_l$ indicates that the density distribution $P_l(\rho_i)$ moves from a Gaussian shape, centered at $\rho_i=\overline{\rho}$, to a binodal shape, where the two maxima of the distribution move progressively towards the coexistence densities $\rho_-$ and $\rho_+$. We show such evolution for both experiments and simulations in Fig.~\ref{fig:distribution}(a-d) for subcells of length $l\simeq3\sigma$.  To be consistent with the phase diagram in Fig.~\ref{fig:phase}, we choose to plot $P_l(\phi)$ instead of $P_l(\rho)$. We start with a hard-sphere sample [Fig.~\ref{fig:distribution}(a)] for which we observe a narrow distribution lacking low ($\phi<0.1$) and high density ($\phi>0.5$) regions. The same behavior continues up to $V=2$ [Fig.~\ref{fig:distribution}(b)]. When further increasing $V$, we start to observe broader distributions with almost empty ($\phi<0.1$) and colloid-rich regions ($\phi>0.5$). This can be clearly seen for our last ergodic sample at $V=2.9$ [Fig.~\ref{fig:distribution}(c)]. Interestingly, dense regions can reach packing fractions larger than the freezing point of a hard-sphere fluid. This explains why for non-polydisperse samples, one can expect a speed up of crystallization around the critical point, where the dense regions arising from criticality will lower the nucleation barrier~\cite{galkin2000control,fortini2008crystallization,savage2009,taylor2012temperature,haxton2015crystallization}. Continuing to quench, the distribution becomes broader and non-Gaussian at $V=3.2$, which also corresponds to the location of our first gel sample. Surprisingly, we find that $P_l^{\text{exp}}$ matches quite closely the equilibrated $P_l^{\text{sim}}$, indicating that aging effects are not pronounced. Finally, we can say that the overall behavior of $P_l(\phi)$ confirms quantitatively the observation made for confocal images in the previous section [Fig.~\ref{fig:phase}(a-d)].

\subsection{Fractal dimension, bulk correlation length, and bond distribution}

\begin{figure*}[t]
  \includegraphics[scale=1.0]{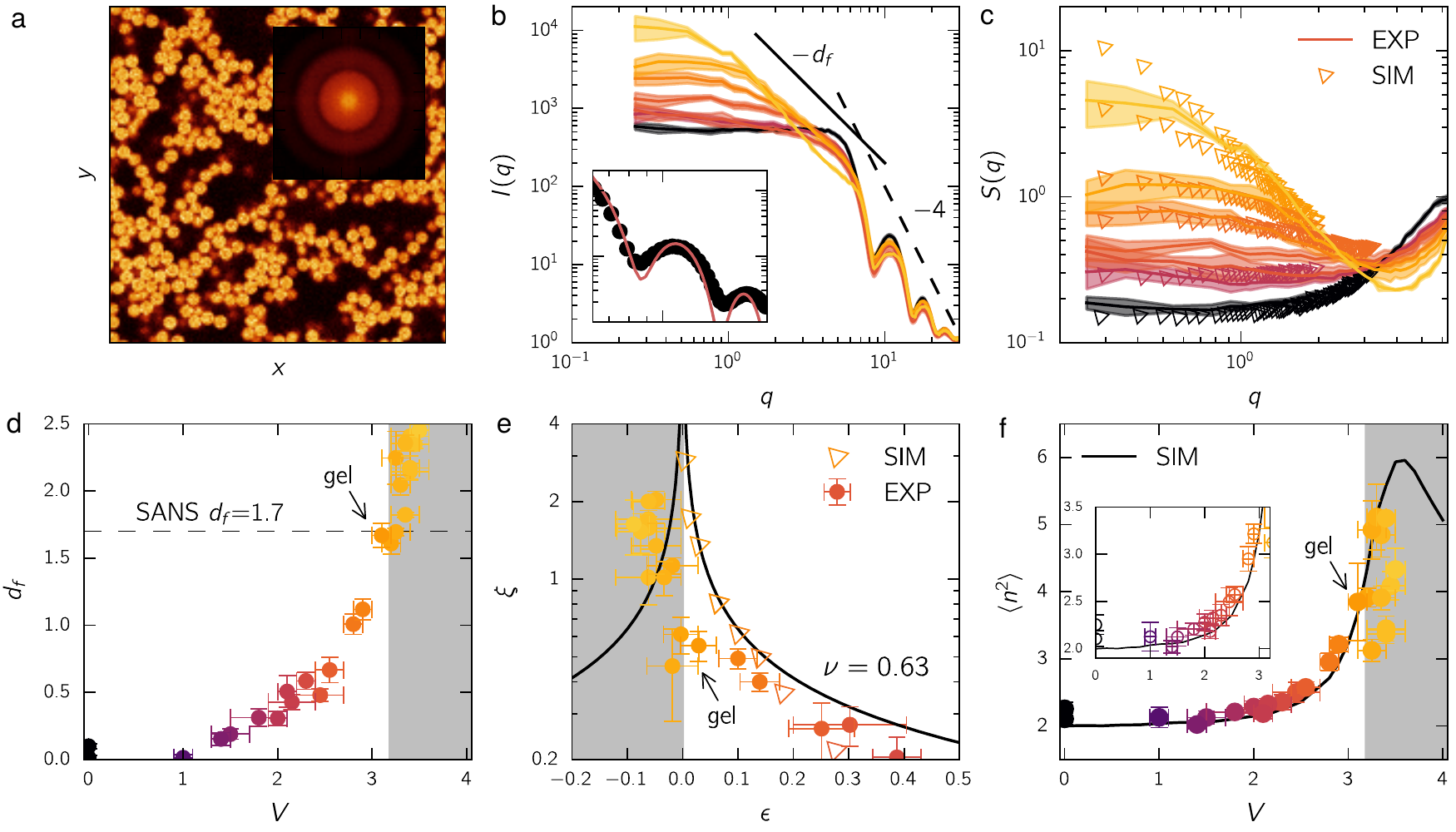}
  \caption{\textbf{Evolution of the Fractal dimension, bulk correlation length, and bond distribution.} (a)~Confocal image of a gel close to criticality. The inset shows the associated scattering pattern. (b)~Scattering intensity $I(q)$ as a function of the wave vector $q$ for various samples along the critical isochore $\phi\simeq\phi_c$. Inset: Fit of the form factor to extract $S(q)$. (c)~Structure factor $S(q)$  as a function of the wave vector $q$ for the same samples. Solid lines are from simulation data at $\phi\simeq\phi_c$. (d) Evolution of the fractal dimension $d_f$ as a function of the attraction strength $V$. The horizontal dashed line indicates the fractal dimension reported for SANS experiments~\cite{eberle2011dynamical}. (e)~Evolution of the bulk correlation length $\xi$ as a function of the reduced attraction strength $\epsilon$. Circles and square are experimental and simulation data, respectively. The black solid line indicates the 3D Ising scaling with $\xi\sim\epsilon^{-0.63}$. (f)~Variance $\langle n^2 \rangle$ of the bond distribution as a function of the attraction strength $V$. The solid black lines indicate the simulation data for the critical isochore. Inset: ergodic fluids before gelation. Arrows indicate the sample where gelation occurs first.}
  \label{fig:scat}
\end{figure*}

We now turn to discuss the characteristic cluster shape and length scale formed by colloidal particles when quenched though criticality. We first compute the scattering intensity $I(q)$ as a function of the wave vector $q$ from the Fourier transform of the confocal pixel map. In Fig.~\ref{fig:scat}(a) we show a typical colloidal gel close to criticality and its associated Fourier spectrum in the inset. A radial average leads to $I(q)$ as show in Fig (b) for different samples along the critical isochore. When quenching, we observe a divergence of $I(q)$ for small $q$ as a consequence of larger domains forming and diffusing light. Since the scattering intensity $I(q)$ is proportional to $P(q)S(q)$, where $P(q)$ is the form factor and $S(q)$ the structure factor. We expect for $q>2\pi/\sigma$ the scaling $I(q)\sim q^{-4}$, which is confirmed by the dashed line in Fig.~\ref{fig:scat}(b). For small $q$, the scaling depends on the shape of the diffusing clusters through $d_f$, its the fractal dimension. Previous small-angle neutron scattering (SANS) experiments have shown that the fractal dimension at the gelation is typically around $d_f\simeq1.7$~\cite{eberle2011dynamical}, which is supported by particle-resolved experiments~\cite{ohtsuka2008,rice2012} and simulation~\cite{griffiths2017local}.

We will discuss this point in detail later on. We can additionally extract $S(q)$ for small wave vector from $S(q)=I(q)/(AP(q))$, where we evaluate $AP(q))$ by fitting $I(q)$ for $q>2\pi/\sigma$ by a polydisperse form factor and a prefactor $A$, as shown in the inset of Fig.~\ref{fig:scat}(b). The resulting procedure is shown in Fig (c) and directly compared to simulation data. We find an overall very good agreement between experimental data and simulations. The structure factor can then be used to extract an estimation of the bulk correlation length $\xi$ through the Orstein-Zernike scaling,
\begin{equation}
  S(q)=\frac{S_0}{1+(\xi q)^2},
\end{equation}
which holds close to criticality for small $q$. In Fig.~\ref{fig:scat}(d), we plot the evolution of the fractal dimension $d_f$ as a function of the attraction strength $V$. We find when increasing $V$ a progressive increase of $d_f$ and saturates for gel samples at $d_f=2.4-2.5$. These values agree with fractal dimensions found for gel at low volume fractions and large attraction strength~\cite{griffiths2017local}. At criticality, where gelation occurs, we find $d_f=1.6-1.7$, which is consistent with a previous SANS experimental study~\cite{eberle2011dynamical}.

In Fig.~\ref{fig:scat}(e), we present the behavior of the bulk correlation length $\xi$ as a function of the reduced attraction strength $\epsilon$. We find for simulation data that $\xi$ diverges for $\xi\to0$ and that values are well modelled by the Ising universality class where $\xi\sim\epsilon^{-\nu}$ up to $\epsilon\simeq0.2$, where in 3D $\nu\simeq0.63$. In contrast, for the experimental data the equilibrium fluids follow closely the simulation data until gelation occurs close to the binodal, whereby the system is arrested and therefore the correlation length does not increase beyond $\xi\approx2\sigma$. This rather small spatial correlation is not in contradiction with confocal images of gels, where structures are quite ramified in a network without large colloidal domains, cf. Fig.~\ref{fig:phase}(e) and Fig.~\ref{fig:scat}(a). This behavior would indicates that gels are, in a way, pictures of early critical fluids, where the cost of breaking bonds does prevent relaxation and thus the growth of the correlation length $\xi$.

Finally, in Fig.~\ref{fig:scat}(d) we discuss the behavior of the bond distribution through its variance $\langle n^2\rangle$. It was shown recently for a similar colloid-polymer mixture model based on the Asakura-Oosawa potential that the variance of the bond distribution as a function of $V$ is peaked close to gelation~\cite{kohl2016directed}. We observe the same behavior in our simulations crossing the binodal, where $\langle n^2\rangle$ exhibits a maximum for $V\simeq3.5$. Quenching further the variance decreases, indicating a progressive aging of the network structure. The behavior of the variance will of course depend on time and other factors such as hydrodynamic interactions~\cite{royall2015probing,varga2016hydrodynamic}. In experiments, we also observe a growth of $\langle n^2\rangle$ for equilibrium fluids ($V<3$). We find larger variances for gel samples, but not as high as in simulations, which also confirms the picture found for $\xi$ in Fig.~\ref{fig:scat}(e), where the gel structure corresponds to an arrested critical fluid. We return to the behaviour in the gel state in Sec.~\ref{enRouteToTheGel}.

\subsection{Local structure}
\label{lfs}

\begin{figure*}[t]
  \includegraphics[width=\linewidth]{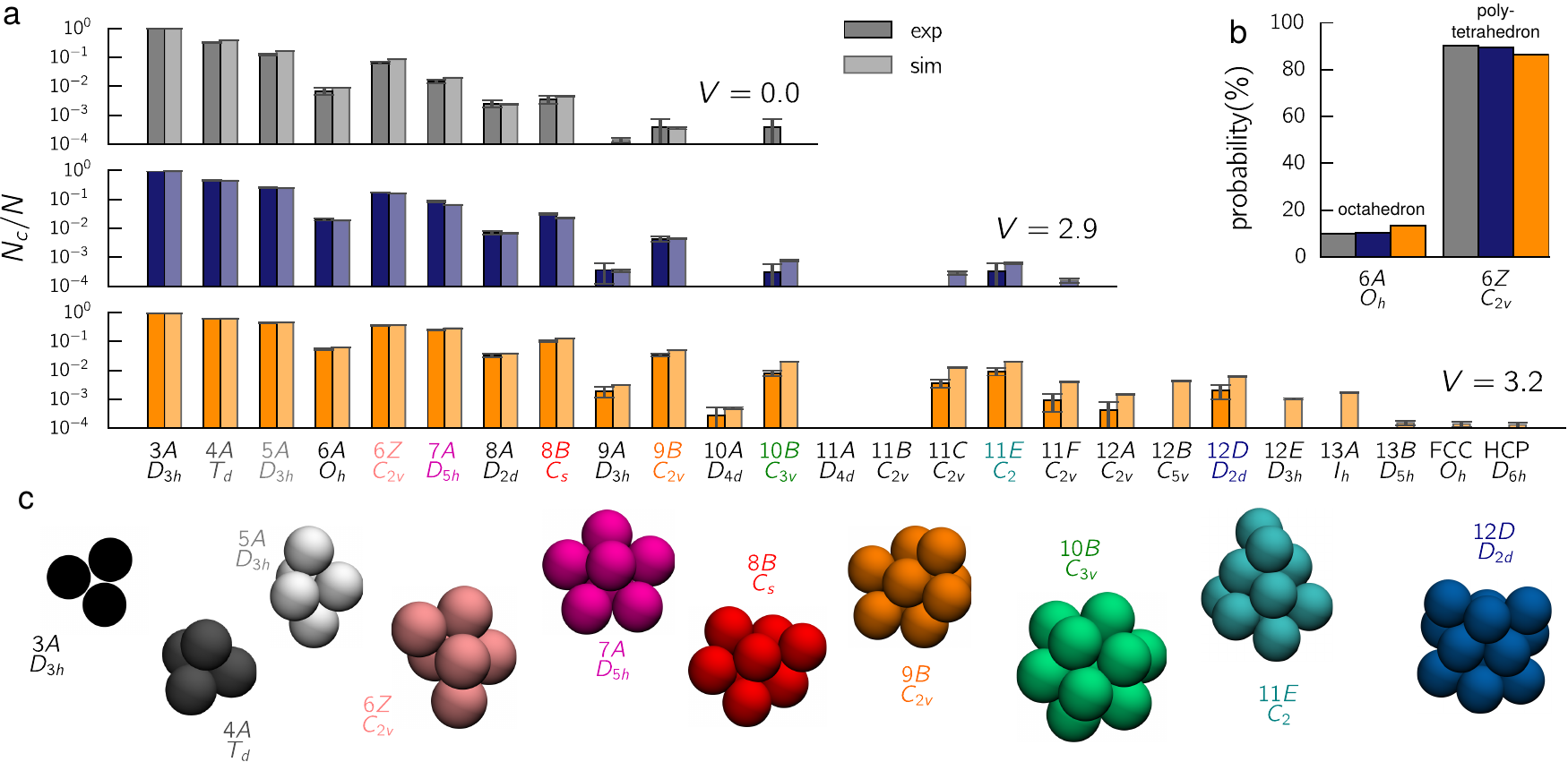}
  \caption{\textbf{Identification of locally favored structures.}~(a) Populations of local clusters for three different samples along the critical density, $\phi_c\simeq0.275$. From top to bottom: a hard-spheres sample, a critical fluid, and an arrested gel. (b) Experimental probability to observe octahedron and poly-tetrahedron in clusters composed of 6 particles. Colors distinguish the three different samples along the critical isochore. (c) Snapshots of selected locally favored structures.}
  \label{fig:tcclfs}
\end{figure*}

We now go to smaller scales and inquire what type of local structures emerge in a fluid close to criticality, and on their role in the dynamical arrest of the gel network. We have seen that in our experiments the typical length scale before gelation reaches $\approx2\sigma$, emphasizing that the change in the dynamics at gelation is quite local and might only be caused by the first neighbor shell of each particle. To gain insight into the local structure of the super-critical fluid, we employ the topological cluster classification (TCC)~\cite{malins2013identification}. We use the same bond network as described in the methods section. From the TCC, we extract the population of clusters of size $m$ composed of $3$ to $13$ particles. This includes crystal structures like face-centered cubic ($\text{fcc}$) and hexagonal close-packed ($\text{hcp}$). In the TCC, each cluster type is labeled with two characters $mX$, with $m$ the number of particles composing the cluster and $X$ a letter to distinguish different spatial symmetries. Those letters follow the convention of the minimum energies of various potentials such as the Morse potential~\cite{doye1995effect}.  We then identify for each cluster size $m$ the predominant type $X$ which corresponds to a minimum energy structure, a locally favored structure (LFS)~\cite{royall2008direct}. In practice, we determine these clusters by following the global population spectrum of all clusters at different values of the attraction strength $V$. The population frequency of a cluster of type $c$ is computed as $N_c/N$, with $N_c$ the number of particles composing a cluster of type $c$ and $N$ the total number of particles. In the experimental data, we do not consider colloidal particles that are less than $\sigma$ away from an edge.

In Fig.~\ref{fig:tcclfs}(a), we show the population of clusters comparing experiments and simulations. We find an excellent agreement, which demonstrates that mapping to the SW fluid reproduces also the higher-order structural features present in the experiments. We pick three different state points along the critical isochore including a hard-sphere fluid ($V=0$), near-critical fluid ($V=2.9$), and gel ($V=3.2$). We first notice that the overall histograms show that when increasing $V$ (from top to bottom), larger clusters composed of more than 10 particles appear. Additionally, we do not observe any signature of crystallization, \emph{i.e.} no fcc nor hcp. However, we find that medium size clusters with $5<m<10$ grow progressively with respect to $V$, and always with the same order among clusters with equal number $m$ of particles. In the following, we exclude larger clusters with $m>12$ from the discussion since their weights are negligible.

\section{Discussion}
\label{discussion}

\subsection{Entropy favors low-symmetry clusters}

For isolated clusters it has been shown that the relative population of clusters with the same number of bonds is determined by entropy~\cite{malins2009,meng2010free}. Interestingly, this still holds for the fluid and even the gel as shown in Fig.~\ref{fig:tcclfs}(b) for $m=6$. We find that $6Z$ clusters, which are polytetrahedra, are always significantly more prevalent in the system compared to $6A$ clusters, which are octahedrons (the population goes up from $\simeq5\%$ for isolated $6A$ to $\simeq10\%$ in the fluid). The same trend can be observed for other clusters, with the more symmetric clusters being less populated. This demonstrates that the minimum (free) energy clusters are determined by rotational and vibrational entropies. It has also been shown that the potential part of the free energy will promote both octahedral and tetrahedral order. The same observation was found in a more recent numerical study of the SW fluid~\cite{haxton2015crystallization}, where gelation was associated with polytetrahedral order. Another important observation is that we also find a large population of clusters with fivefold symmetry ($8B$, $9B$, and $10B$) at any $V$. These structures are known to be local energy minima of the Morse potential~\cite{doye1995effect}. Hence, even though a gel is in an energy landscape far away from the equilibrium state, it can still locally minimize its free energy by forming isolated locally favored structures, which eventually will overlap and form the gel network. For larger clusters with $11$ and $12$ particles, we observe several Morse minima that correspond to $10<\rho_0<25$ ($\rho_0$ is the Morse potential parameter controlling the range of attraction, in our case $\delta$). The range for $\rho_0$ found here is consistent with short range attraction. We found that $11E$ and $12D$ are the dominant structures for $m=11$ and $m=12$ respectively, which corresponds to $\rho_0\simeq17$~\cite{doye1995effect}.

\subsection{Hard sphere fluid}

\begin{figure}[t]
  \includegraphics[scale=1.0]{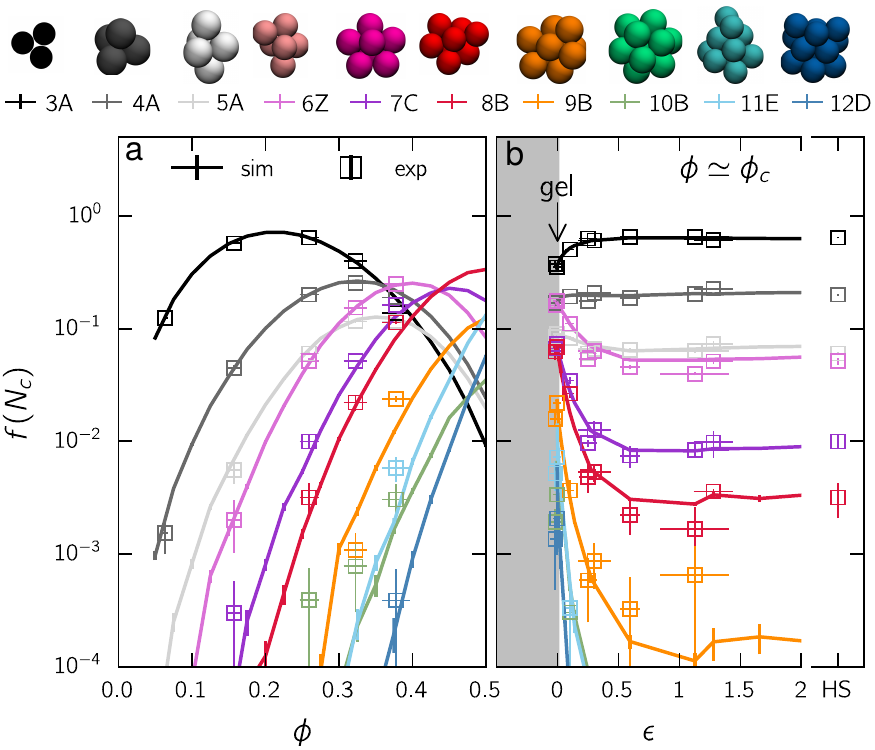}
  \caption{\textbf{Evolution of LFS through compression and cooling.} (a)~Evolution of the fraction of locally favored structures as a function of the packing fraction $\phi$ for hard spheres ($V=0$). (b)~Evolution of the fraction of locally favored structures as a function of the reduced attraction strength $\epsilon$. The yellow area indicates the two-phase region where gelation occurs. Solid lines and open squares indicate results for simulations and experiments, respectively.}
  \label{fig:lfs}
\end{figure}

\begin{figure}[t]
  \includegraphics[scale=1.0]{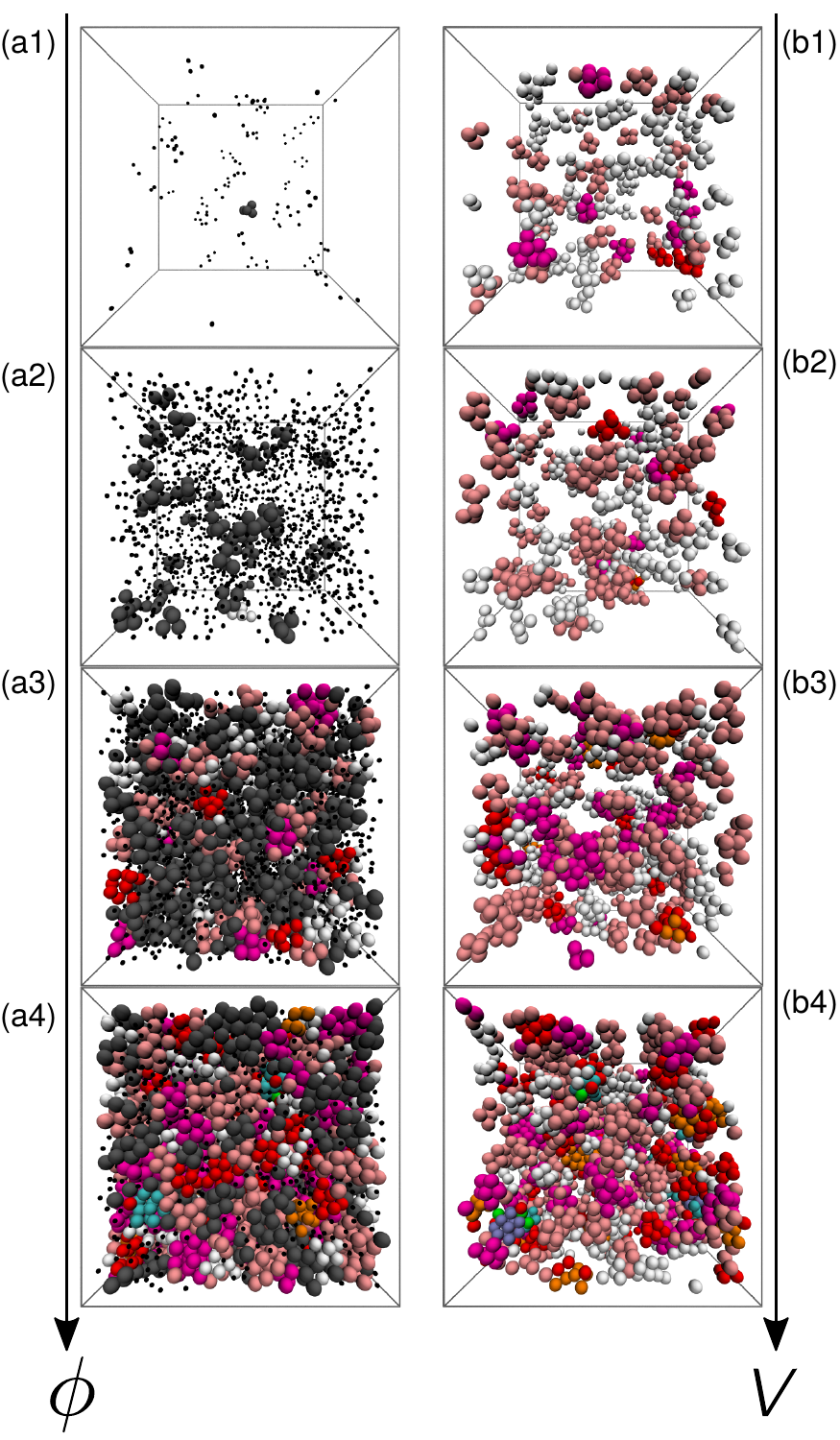}
  \caption{\textbf{Experimental snapshots of the cluster evolution.}~ Evolution of the clusters when increasing the packing fraction $\phi$ (left) and attraction strength $V$ (right). Only structures with $m>4$ are shown for the cooling path.}
  \label{fig:snapshots}
\end{figure}

Having obtained a set of minimum energy clusters, we use these to follow the overall structural change in the fluid. As a reference, we first consider the compression of the hard sphere fluid ($V=0$) towards the glass shown in Fig.~\ref{fig:lfs}(a). At low density, $\phi\simeq0.05$, there is an absence of clusters as one can expect since particles are mostly isolated without any neighbors. At this low density, the system exhibits as a larger cluster only a few percents of $4A$ (tetrahedron). The typical spatial arrangement of clusters is illustrated with an experimental snapshot in Fig.~\ref{fig:snapshots}(a1). We then observe a rapid change from $\phi=0.05$ to $0.25$, where medium sized clusters such as $5A$, $6Z$, $7C$, and $8B$ start to appear, cf. Fig.~\ref{fig:snapshots}(a1-a3). This continues until a crossover at $\phi>0.25$, where isolated and small clusters ($3A$ and $4A$) are converted with a combination of medium clusters ($5A$, $6Z$, $7C$, and $8B$) to larger structures: $9B$, $10B$, $11E$, and $12D$, see Fig.~\ref{fig:snapshots}(a4). This kind of conversion will continue at higher packing fractions with the extinction of $5A$, $6Z$, and $7C$ to promote clusters sharing the same sub-structures. We also notice that the agreement between experiments and simulation is excellent. This gives another confirmation for the robustness of the mapping procedure employed.

\subsection{En route to the gel}
\label{enRouteToTheGel}

In Fig.~\ref{fig:lfs}(b), we follow the same idea but now approaching the gel. We fix the density to the critical packing fraction $\phi_c=0.275$ and progressively increase the attraction strength $V$ towards the binodal. We find a plateau with little change in structure until $V\simeq2$. We show the spatial arrangement of clusters at $V=1.1$ in Fig.~\ref{fig:snapshots}(b1), where we only display clusters with $m>4$. For higher attraction strengths, $V>2$, we observe a structural crossover (cf. the compression of hard spheres). Clusters with $m>4$ increase quickly until reaching the binodal. Interestingly, this change starts where we have located the beginning of the critical scaling for the bulk correlation length $\xi$ and the static susceptibility $\chi$. The main contribution to this structural evolution comes from the creation of a large amount of $6Z$, $7C$, and $8B$ clusters, see Fig.~\ref{fig:snapshots}(b1-b3). These clusters are not necessarily localized in space, but we do observe denser regions with even the presence of $9B$ at $V=2.9$, see Fig.~\ref{fig:snapshots}(b3). This is consistent with our previous finding in Fig.~\ref{fig:distribution}(c), where the density distribution exhibits a broader tail toward large packing fractions $\phi>0.4$. Therefore, we can expect a link between the local structures of dense hard spheres and the colloidal-rich regions arising from critical density fluctuations. Finally at $V>V_c$, which corresponds to our first identification of a gel, clusters percolate the whole system, see Fig.~\ref{fig:snapshots}(b4). The inner parts of the percolating structure are rich in large clusters such as $8B$, $9B$, and $10B$, which are defective icosahedra sharing a fivefold symmetry. They are known to play an important role in the slowing down of the dynamics for dense hard-spheres approaching the glass transition~\cite{karayiannis2011fivefold,taffs2013structure,pinchaipat2017}. This may help to explain the origin of the rigidity of the network leading to gelation~\cite{royall2018vitri,lu2008gelation}.

The crucial difference between a gel and a dense hard sphere liquid is that not only the defective icosahedra ($8B$, $9B$, and $10B$) are very stable due to entropy, but each bond breaking will also result in an energetic penalty of more than $3k_BT$. Therefore only colloids in the outer part of dense regions will be able to break bonds and diffuse (see Supplementary Video). To conclude, we find that the structural crossover starts $1k_BT$ before the location of the binodal when increasing $V$. This is a direct consequence of criticality inducing larger density fluctuations. For a monodisperse sample, these fluctuations might promote crystal precursors and lower the nucleation time \cite{haxton2015crystallization}. For a polydisperse sample, however, they induce low-symmetry polytetrahedral backbones for the bond network. These clusters are incompatible with respect to the crystal symmetry. Hence, the system will fall into an amorphous solid state, the gel. Regarding the mapping we comment that, since the fluid is ergodic, the mapping to the SW fluid should be robust (as indeed it is). In contrast, the gel is non-ergodic and thus the correspondence would depend on the dynamics and history~\cite{royall2015}.

\section{Conclusions}
\label{conclusion}

We have investigated the role of criticality in the gelation of sticky spheres. Combining experiment and simulations, we provide further evidence that the dynamical arrest is initiated by the onset of critical fluctuations in agreement with previous work~\cite{manley2005glasslike,lu2008gelation,royall2008direct,royall2018vitri}. We have demonstrated that carefully mapping two-point structure to the square-well fluid faithfully reproduces the experimental data including higher-order local structures as identified by the topological cluster classification method~\cite{malins2013identification}. In particular, we find that gel samples are (i)~located at the cumulant crossing, (ii)~identified by a broad distribution of densities, and (iii)~have correlation lengths of $\xi\simeq2\sigma$. We find a sharp but continuous increase of locally favored structures when increasing the attraction strength. This increase occurs in concert with the increase of both the bulk correlation length and the static susceptibility, which can be extracted together with the structure factor of the fluid. More precisely, the start of the critical scaling of these two quantities coincides with the appearence of larger locally favored structures. The picture of a gel is thus that of an early critical fluid, which is arrested due to the large cost of breaking bonds. Before arrest, clusters of several particles appear, which have a low symmetry favored by entropy. The densification of these clusters driven by the incipient critical fluctuations then leads to the gel.

%% ---- acknowledgments ----

\acknowledgements

We gratefully acknowledge Francesco Turci and Matteo Campo for stimulating discussions. We thank M. Schmiedeberg for helpful discussions. We acknowledge financial support by the DFG through collaborative research center TRR 146 (D.R.).

\appendix

\section{Effective colloid diameter}
\label{sigma}

\begin{figure}[t]
  \includegraphics[scale=1.0]{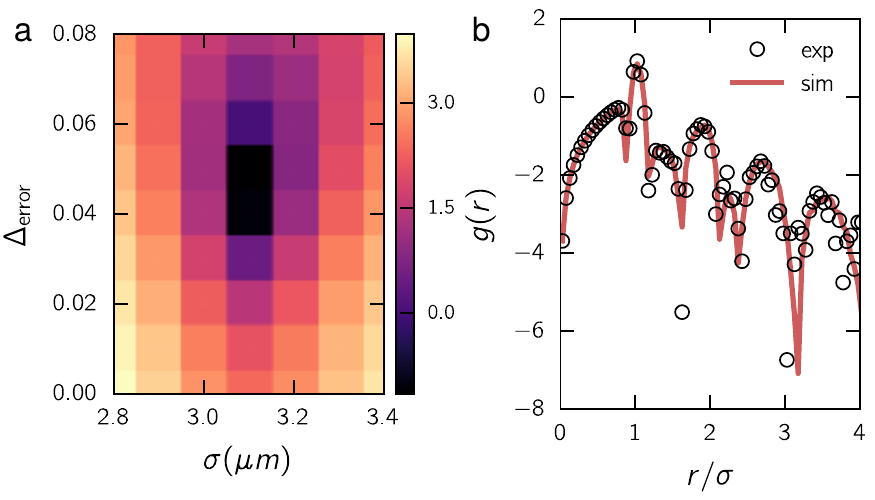}
  \caption{\textbf{Effective diameter and tracking errors.}~(a) Colormap of $\ln(\chi^2)$ against the effective diameter $\sigma$ and the tracking error $\Delta_{\text{error}}$. (b) Resulting matching of the pair correlation between an experimental sample and simulation at criticality ($\phi=0.275,V=3.2$) }
  \label{fig:diameter}
\end{figure}

There are two challenges for the determination of the effective packing fraction and the effective temperature of a colloidal sample from real space imaging. First, it is difficult to know the change of the colloidal diameter in solution due to swelling and unscreened electrostatic~\cite{poon2012measuring,royall2013search}. This leads to a poor estimation of the sample packing fraction and, \emph{e.g.}, a mismatch of structural oscillations in pair correlations. Second, errors due to imaging and tracking will lead to an error on the true position of a particle, resulting in a broadening of the peaks of a pair correlation~\cite{royall2007measuring}. In the context of matching a sample's $g(r)$ onto simulations, these errors will induce a systematic overestimation of the temperature of a mixture at a given polymer concentration $c_p$. To handle these issues, we pick a sample close to criticality and compute its pair correlation $r[g(r)-1]$ for various effective diameters. To mimic tracking errors, we pick a simulation state point at criticality and apply different Gaussian noises of variance $\Delta_{\text{error}}$ to the particle positions. We then find the optimal set of parameters that minimize $\chi^2 = \sum_i [ r_i(h_{\text{exp}}(r_i) - h_{\text{sim}}(r_i))]^2$. The evolution of $\chi^2$ as a function of the parameters is shown in Fig.~\ref{fig:diameter}(a). We find a unique minimum for $\chi^2$ leading to an effective diameter $\sigma=3.1\mu m$ (5\% larger than for dry colloids) and a tracking error $\Delta_{\text{error}}=5\%$, which is consistent with previous works~\cite{royall2007measuring}. The resulting matched pair correlation function is shown in Fig.~\ref{fig:diameter}(b). We observe a very good agreement between experiment and simulation which validate $\sigma$ and $\Delta_{\text{error}}$.

%% ---- bibliography ----

\end{document}